\documentclass[envcountsame]{llncs}
\usepackage{graphicx}
\usepackage{amsmath,amssymb}
\usepackage{revsymb}
\usepackage{color}
\usepackage{bbm}
\usepackage{hyperref}
\usepackage{theorem}
\usepackage{latexsym}

\newlength{\blank}
\newcommand{\ket}[1]{|#1\rangle}
\newcommand{\bra}[1]{\langle#1|}

\mathchardef\ordinarycolon\mathcode`\:
\mathcode`\:=\string"8000
\def\vcentcolon{\mathrel{\mathop\ordinarycolon}}
\begingroup \catcode`\:=\active
  \lowercase{\endgroup
  \let :\vcentcolon
  }

\newcommand{\nc}{\newcommand}
\nc{\rnc}{\renewcommand}
\nc{\beq}{\begin{equation}}
\nc{\eeq}{{\end{equation}}}
\nc{\beqa}{\begin{eqnarray}}
\nc{\eeqa}{\end{eqnarray}}
\nc{\lbar}[1]{\overline{#1}}
\nc{\ketbra}[2]{|#1\rangle\!\langle#2|}
\nc{\proj}[1]{| #1\rangle\!\langle #1 |}
\nc{\avg}[1]{\langle#1\rangle}
\nc{\Rank}{\operatorname{Rank}}
\nc{\smfrac}[2]{\mbox{$\frac{#1}{#2}$}}
\nc{\tr}{\operatorname{Tr}}
\nc{\ox}{\otimes}
\nc{\dg}{\dagger}
\nc{\dn}{\downarrow}
\nc{\cA}{\mathcal{A}}
\nc{\cB}{\mathcal{B}}
\nc{\cC}{\mathcal{C}}
\nc{\cD}{\mathcal{D}}
\nc{\cE}{\mathcal{E}}
\nc{\cF}{\mathcal{F}}
\nc{\cG}{\mathcal{G}}
\nc{\cH}{\mathcal{H}}
\nc{\cI}{\mathcal{I}}
\nc{\cJ}{\mathcal{J}}
\nc{\cK}{\mathcal{K}}
\nc{\cL}{\mathcal{L}}
\nc{\cM}{\mathcal{M}}
\nc{\cN}{\mathcal{N}}
\nc{\cO}{\mathcal{O}}
\nc{\cP}{\mathcal{P}}
\nc{\cR}{\mathcal{R}}
\nc{\cS}{\mathcal{S}}
\nc{\cT}{\mathcal{T}}
\nc{\cX}{\mathcal{X}}
\nc{\cY}{\mathcal{Y}}
\nc{\cZ}{\mathcal{Z}}
\nc{\csupp}{{\operatorname{csupp}}}
\nc{\qsupp}{{\operatorname{qsupp}}}
\nc{\var}{\operatorname{var}}
\nc{\rar}{\rightarrow}
\nc{\lrar}{\longrightarrow}
\nc{\polylog}{\operatorname{polylog}}
\nc{\1}{{\bbbone}}
\nc{\id}{{\operatorname{id}}}

\nc{\RR}{{{\mathbb R}}}
\nc{\CC}{{{\mathbb C}}}
\nc{\FF}{{{\mathbb F}}}
\nc{\NN}{{{\mathbb N}}}
\nc{\ZZ}{{{\mathbb Z}}}
\nc{\PP}{{{\mathbb P}}}
\nc{\QQ}{{{\mathbb Q}}}
\nc{\UU}{{{\mathbb U}}}
\nc{\EE}{{{\mathbb E}}}

\nc{\QED}{{$\hfill\Box$}}

\newcommand\blfootnote[1]{%
  \begingroup
  \renewcommand\thefootnote{}\footnote{#1}%
  \addtocounter{footnote}{-1}%
  \endgroup
}

\begin{document}

\pagestyle{headings} 

\title{Identification Via Quantum Channels\,${}^\ast$}
\titlerunning{Identification Via Quantum Channels}

\author{Andreas Winter}
\authorrunning{Andreas Winter}

\institute{ICREA -- Instituci\'{o} Catalana de Recerca i Estudis Avan\c{c}ats\protect\\
Pg. Lluis Companys 23, ES-08010 Barcelona, Spain\protect\\[1mm]
F\'{\i}sica Te\`{o}rica: Informaci\'{o} i Fenomens Qu\`{a}ntics\protect\\
Universitat Aut\`{o}noma de Barcelona, ES-08193 Bellaterra (Barcelona), Spain\protect\\[2mm]
Department of Mathematics, University of Bristol, Bristol BS8 1TW, U.K.\protect\\[2mm]
Centre for Quantum Technologies, National University of Singapore\protect\\
2 Science Drive 3, Singapore 117542, Singapore\protect\\[2mm]
%
}


\maketitle

\begin{abstract}
  We review the development of the quantum version of Ahls\-wede and
  Dueck's theory of identification via channels. As is often the case
  in quantum probability, there is not just one but several quantizations:
  we know at least two different concepts of identification of classical
  information via quantum channels, and three different identification
  capacities for quantum information.
  
  In the present summary overview we concentrate on conceptual points
  and open problems, referring the reader to the small set of original
  articles for details.
\end{abstract}

\begin{center}
  \emph{\small Dem Andenken an Rudolf Ahlswede (15/9/1938---18/12/2010)}
\end{center}

\blfootnote{${}^\ast$\,Original date 30 November 2012; \textcolor{blue}{annotated erratum 14 February 2024}.}

\vspace{0.707cm}

\setcounter{section}{-1}

\section{Quantum and classical channels}
\label{sec:channels}
Our communication model is the quantum channel, also known as
completely positive and trace preserving (cptp) linear map between
quantum systems,
\[
  \cN : \cL(A) \longrightarrow \cL(B).
\]
Here, as in the rest of the paper, we assume that $A$, $B$, etc, are
finite dimensional (complex) Hilbert spaces and $\cL(A)$ is the
set of linear operators (matrices) over $A$.

The cptp condition is necessary and sufficient for $\cN$ mapping
states on $A$, i.e.~density operators $\rho \geq 0$ with $\tr\rho=1$, whose
set we denote as $\cS(A)$, to states on $B$, 
and the same for $\cN \ox \id_C$ for arbitrary systems $C$.
Thus, the class of cptp maps is closed under composition, tensor products 
and taking convex combinations. One of the most useful characterizations of 
cptp maps is in terms of the Stinespring dilation~\cite{Stinespring}: namely,
$\cN$ is cptp if and only if there exists an ancilla (environment) 
system $E$ and an isometry $V:A\hookrightarrow B\ox E$ such that
$\cN(\rho) = \tr_E V\rho V^\dagger$. The isometry $V$
is essentially unique, up to unitary equivalence of $E$; hence it makes
sense to define, for a chosen dilation $V$, the \emph{complementary channel}
\[
  \widehat{\cN}: \cL(A) \longrightarrow \cL(E),
\]
by $\widehat{\cN}(\rho) := \tr_B V\rho V^\dagger$.

For a given channel $\cN$, we are interested in the asymptotic 
performance of many iid copies, $\cN^{\ox n}$. 
One can also consider more complicated channel models (such as
with feedback, or with pre-shared correlations), but here we will restrict
ourselves to the simple forward channel -- see however~\cite{winter:q-ID-2}
and~\cite{remote-state-preparation}.

Classical channels are of course transition probability kernels
$N:\cX \rightarrow \cY$ (with finite input and output alphabets $\cX$
and $\cY$, respectively). Such a channel may be identified with the
cptp map
\begin{align*}
  \cN:\cL(\CC^{\cX}) &\longrightarrow \cL(\CC^{\cY}) \\
                \rho &\longmapsto \sum_{xy} N(y|x) \ketbra{y}{x} \rho \ketbra{x}{y},
\end{align*}
while a probability distribution $P$ on $\cX$ is identified with the
state $\sum_x P(x) \proj{x}$.

Two special classes of channels we will have occasion to consider
are the following, either whose input or whose output is classical:
A \emph{cq-channel} $\cN:\cX \longrightarrow \cS(B)$ is a cptp map of 
the form
\[
  \cN(\xi) = \sum_x \bra{x} \xi \ket{x} \rho_x,
\]
with states $\rho_x$ on $B$.
A \emph{qc-channel} $\cM:\cS(A) \longrightarrow \cY$ instead is given
by a quantum measurement, i.e.~a positive operator values measure
(POVM) $(M_y)_{y\in\cY}$ such that $M_y \geq 0$ and $\sum_y M_y = \1$.
The channel then has the form
\[
  \cM(\rho) = \sum_y \tr \rho M_y \proj{y}.
\]

We refer the reader to the excellent text~\cite{Wilde:book} for more
details on quantum and classical channels, and the various transmission
capacities associated with them, including their history.
Here we need only two, the classical and the quantum capacity of a 
channel, $C(\cN)$ and $Q(\cN)$, respectively,
defined as the maximum rates of asymptotically faithful transmission of
classical bits and qubits, respectively, over many iid copies of the channel.
They can be expressed as regularizations of entropic information quantities, 
based on the von Neumann entropy $S(\rho) = -\tr \rho \log \rho$ of a quantum
state $\rho$.
They are given by the formulas
\begin{equation}
\label{eq:C-cap}
\begin{split}
  C(\cN)       &= \lim_{n\rightarrow\infty} \frac{1}{n} C^{(1)}(\cN^{\ox n}), \text{ with} \\
  C^{(1)}(\cN) &= \max_{\{p_x,\rho_x\}} 
                    S\left( \sum_x p_x \cN(\rho_x) \right) - \sum_x p_x S\bigl( \cN(\rho_x) \bigr),
\end{split}\end{equation}
and
\begin{equation}
\label{eq:Q-cap}
\begin{split}
  Q(\cN)       &= \lim_{n\rightarrow\infty} \frac{1}{n} Q^{(1)}(\cN^{\ox n}), \text{ with} \\
  Q^{(1)}(\cN) &= \max_{\rho \in \cS(A)} 
                    S\bigl( \cN(\rho) \bigr) - S\bigl( \widehat{\cN}(\rho) \bigr),
\end{split}\end{equation}
both of which represent the culmination of concerted efforts of several
researchers in the 1990s and early 2000s
(Holevo-Schumacher-Westmoreland
and 
Schumacher \&{} Lloyd-Shor-Devetak, 
respectively). The classical capacity generalizes
Shannon's channel capacity for classical channels $N$, for which $C(N) = C^{(1)}(N)$
reduces to the famous formula in terms of the mutual information~\cite{Shannon}.
%

\medskip
The structure of the rest of the paper is as follows: In section~\ref{sec:ID} we
present the definitions for identification of classical information
via quantum channels, after L\"ober~\cite{Loeber:PhD}, generalizing the 
model of Ahlswede and Dueck~\cite{AhlswedeDueck:ID.A,AhlswedeDueck:ID.B}. 
In section~\ref{sec:q-ID} we move to identification of quantum 
information; section~\ref{sec:weak-decoupling} presents the recently
developed theoretical underpinning to prove capacity formulas for
two of the three quantum models.
In section~\ref{sec:Q-to-C} we show how the quantum identification
results imply new lower bounds on classical identification capacities,
which we illustrate with several examples, shedding new light also
on L\"ober's founding work~\cite{Loeber:PhD}.
Finally, section~\ref{sec:conclusion} is devoted to an outlook on
open questions and possible conjectures.

\section{Classical Identification}
\label{sec:ID}
Ahlswede and Dueck~\cite{AhlswedeDueck:ID.A,AhlswedeDueck:ID.B} introduced
identification by noting that while Shannon's theory of transmission presumes
that the receiver wants to know everything about the message, in reality
he may be interested only in certain aspects of it. In other words, the
receiver may want to compute a function of the message. The most
extreme case is that of identification: for sent message $m$ and an
arbitrary message $m'$, the receiver would like to be able to answer
the question ``Is $m=m'$?'' as accurately as possible.

\begin{definition}[L\"ober~\cite{Loeber:PhD}]
  \label{defi:c-ID-code}
  A \emph{classical identification code for the channel}
  $\cN$ with \emph{error probability $\lambda_1$
  of first, and $\lambda_2$ of second kind} is a set
  $\{(\rho_i,D_i):i=1,\ldots,N\}$ of states $\rho_i$ on $A$ and
  operators $D_i$ on $B$ with $0\leq D_i\leq \1$, i.e.~the pair
  $(D_i,\1-D_i)$ forms a measurement, such that
  \begin{align*}
    \forall i       &\quad \tr\bigl( \cN(\rho_i) D_i \bigr) \geq 1-\lambda_1, \\
    \forall i\neq j &\quad \tr\bigl( \cN(\rho_i) D_j \bigr) \leq \lambda_2.
  \end{align*}
  For the special case of memoryless channels $\cN^{\otimes n}$, 
  we speak of an \emph{$(n,\lambda_1,\lambda_2)$-ID code}, and denote
  the largest size $N$ of such a code $N(n,\lambda_1,\lambda_2)$.

  An identification code as above is called \emph{simultaneous} if
  all the $D_i$ are coexistent: this means that there exists a positive
  operator valued measure (POVM) $(E_t)_{t=1}^T$ and \textcolor{blue}{subsets}
  ${\cal D}_i\subset\{1,\ldots,T\}$ such that $D_i=\sum_{t\in{\cal D}_i} E_t$.
  The largest size of a simultaneous $(n,\lambda_1,\lambda_2)$-ID code
  is denoted $N_{\rm sim}(n,\lambda_1,\lambda_2)$.
  
  Note that $N_{\rm sim}(n,\lambda_1,\lambda_2) = N(n,\lambda_1,\lambda_2) = \infty$
  if $\lambda_1+\lambda_2 \geq 1$, hence to avoid this triviality one
  has to assume $\lambda_1+\lambda_2 < 1$.
\end{definition}

It is straightforward to verify that in 
the case of a classical channel, this definition reduces to
the famous one of Ahlswede and Dueck~\cite{AhlswedeDueck:ID.A}, in particular
all codes are without loss of generality automatically simultaneous. It was
in fact L\"ober~\cite{Loeber:PhD} in his PhD thesis who noticed that in the
quantum case we have to make a choice -- whether the receiver should be
able to answer \emph{all} or \emph{any one} of the ``Is the message $=m'$?''
questions. It was the original realization of Ahlswede and Dueck~\cite{AhlswedeDueck:ID.A}
that $N(n,\lambda_1,\lambda_2)$ grows doubly exponential in $n$, hence the
following definition of the (classical) identification capacity:

\begin{definition}
  \label{defi:ID-capacities}
  The \emph{(simultaneous) classical ID-capacity} of a quantum channel $\cN$
  is given by
  \begin{align*}
    C_{\rm ID}(\cN) &= \inf_{\lambda > 0} 
                       \liminf_{n\rightarrow \infty} \frac{1}{n} \log\log N(n,\lambda,\lambda), \\
    C_{\rm ID}^{\rm sim}(\cN) &= \inf_{\lambda > 0} 
                       \liminf_{n\rightarrow \infty} \frac{1}{n} \log\log N_{\rm sim}(n,\lambda,\lambda),
  \end{align*}
  respectively. 
  We say that the \emph{strong converse} holds for the identification capacity
  if for all $\lambda_1+\lambda_2 < 1$,
  \[
    \lim_{n\rightarrow \infty} \frac{1}{n} \log\log N(n,\lambda_1,\lambda_2) = C_{\rm ID}(\cN),
  \]
  and similarly for $ C_{\rm ID}^{\rm sim}$.
\end{definition}

\begin{theorem}[Ahlswede/Dueck~\cite{AhlswedeDueck:ID.A}, 
  Han/Verdu~\cite{HanVerdu:ID-strong-converse,HanVerdu:resolvability},
  Ahlswede~\cite{Ahlswede:ID-strong-converse}]
  \label{thm:classical-ID-capacity}
  For a classical channel $N$ and any $\lambda_1, \lambda_2 > 0$ with
  $\lambda_1+\lambda_2 < 1$,
  \[
    \lim_{n\rightarrow \infty} \frac{1}{n} \log\log N(n,\lambda_1,\lambda_2) = C(N),
  \]
  in particular, $C_{\rm ID}^{\rm sim}(N) = C_{\rm ID}(N) = C(N)$.
  \qed
\end{theorem}

The direct part of the above theorem, due to Ahlswede and Dueck~\cite{AhlswedeDueck:ID.A}, 
can be seen by concatenating a sufficiently good Shannon channel code with an
identification code for the ideal bit channel. For the latter, \cite{AhlswedeDueck:ID.A}
contains a combinatorial construction showing that by $k$-bit encodings, one can
identify $\geq 2^{\Omega(2^k)}$ messages. Using this, the direct part of the following
result is immediate:

\begin{theorem}[L\"ober~\cite{Loeber:PhD}, Ahlswede/Winter~\cite{AhlswedeWinter:ID-q}]
  \label{thm:quantum-channel-ID}
  For quantum channel $\cN$, 
  \[
    C_{\rm ID}(\cN) \geq C_{\rm ID}^{\rm sim}(\cN) \geq C(\cN).
  \]
  The simultaneous ID-capacity obeys a strong converse under the additional
  restriction that the signal states $\rho_i$ are from a set that is the convex
  hull of $\leq 2^{2^{o(n)}}$ quantum states on $A^n$. I.e., denoting the maximum
  number of messages under this constraint by
  $\underline{N}_{\rm sim}(n,\lambda_1,\lambda_2)$,
  \[
    \limsup_{n\rightarrow \infty} 
         \frac{1}{n} \log\log \underline{N}_{\rm sim}(n,\lambda_1,\lambda_2) \leq C(\cN),
  \]
  for $\lambda_1+\lambda_2 < 1$. (For instance, the $\rho_i$ could be restricted
  to be -- approximately -- separable states.)
  
  For cq-channels, the constraint is w.l.o.g.~satisfied since there are
  only $|\cX|^n$ classical input symbols, so for these channels the simultaneous
  ID-capacity obeys a strong converse, with $C_{\rm ID}^{\rm sim}(\cN) = C(\cN) = C^{(1)}(\cN)$.

  Indeed, in the case of cq-channels,
  the strong converse holds even without the simultaneity constraint:
  \[
    \lim_{n\rightarrow \infty} \frac{1}{n} \log\log N(n,\lambda_1,\lambda_2) = C(\cN) = C^{(1)}(\cN),
  \]
  for $\lambda_1+\lambda_2 < 1$.
  \qed
\end{theorem}
[To be precise, L\"ober's results are in the framework of Han and 
Verd\'{u}~\cite{HanVerdu:ID-strong-converse,HanVerdu:resolvability}, of
``arbitrary'' sequences of channels and using information spectrum methods. 
As we are focusing on the iid case here, we stated only a special case
of his theorem.]

The last, non-simultaneous part of the Theorem is the main identification 
result of~\cite{AhlswedeWinter:ID-q}, which was proved by developing a 
theory of tail bounds for sums of random matrices, extending classical
Hoeffding bounds, and inspired by Ahlswede's strong converse for the ID-capacity
of classical channels~\cite{Ahlswede:ID-strong-converse}. The simplest,
and most useful, version is as follows.
\begin{lemma}[Ahlswede/Winter~\cite{AhlswedeWinter:ID-q}]
  \label{lemma:op-Chernoff}
  For i.i.d.~random variables $X_i$ in $d\times d$ Hermitian matrices
  and $0 \leq X_i \leq \1$, such that $\EE X_i = \mu\1$. Then, for
  $\mu \leq \alpha \leq 1$ and $0 \leq \alpha \leq \mu$, respectively,
  \begin{equation*}\begin{split}
    \Pr&\left\{ \frac{1}{n}\sum_{i=1}^n X_i \not\leq \alpha\1 \right\} \leq d\, e^{-n D(\alpha\|\mu)}, \\
    \Pr&\left\{ \frac{1}{n}\sum_{i=1}^n X_i \not\geq \alpha\1 \right\} \leq d\, e^{-n D(\alpha\|\mu)},
  \end{split}\end{equation*}
  with the binary relative entropy 
  $D(\alpha\|\mu) = \alpha\ln\frac{\alpha}{\mu} + (1-\alpha)\ln\frac{1-\alpha}{1-\mu}$.
  
  As a consequence, for all $0 \leq \epsilon \leq \frac12$,
  \[
    \Pr\left\{ \frac{1}{n}\sum_{i=1}^n X_i \not\in [(1-\epsilon)\mu\1,(1+\epsilon)\mu\1] \right\} 
                                                               \leq 2d\, e^{-\frac14 n \mu \epsilon^2}.
  \]
  using elementary estimates for the relative entropy.
  \qed
\end{lemma}
The power of this Lemma is in its giving explicit and simple tail bounds,
useful already for finite $n$ and $d$, whereas general abstract large
deviation theory -- which applies, see~\cite{AhlswedeBlinovsky} for a
version in infinite dimension -- often incurs complex finite $n$ behaviour,
only yielding clear asymptotic statements. The proof of the Lemma is simple,
too: it requires generalizing the elementary Markov-Chebyshev
inequalities and the Bernstein trick from real random variables to matrices.
It has since found countless applications in quantum information
theory and beyond: The first proofs of some core results such as the 
quantum channel capacity, remote state preparation or decoupling 
heavily relied on it, cf.~\cite{Wilde:book}, as did the structurally 
simple proof of the Alon-Roichman theorem and matrix versions of compressed 
sensing, cf.~\cite{Tropp:notes} and references therein. 
The latter also presents far-reaching generalizations of the above bounds. 

\medskip
It is not known whether simultaneous and non-simultaneous ID-capacity
coincide or not for general quantum channels. 
%
In any case, going beyond simultaneity seems to provide major freedom:

\begin{example}
  \label{ex:fingerprinting}
  Buhrman \emph{et al.}~\cite{fingerprinting} found that in the
  space of $n$ qubits, whilst the largest number of orthogonal pure state
  vectors is clearly the dimension of the Hilbert space, $2^n$, there
  are $N \geq 2^{\Omega(2^n)}$ \emph{pairwise almost orthogonal} pure states,
  i.e.~$|\langle \psi_i \ket{\psi_j}| \leq \epsilon$ for $i\neq j$.
  
  They dubbed this ``fingerprinting'' because a verifier who gets a copy of
  each $\ket{\psi_i}$ and $\ket{\psi_j}$ can efficiently determine whether
  $i=j$ or not. In particular, the set of these vectors forms a 
  (non-simultaneous) ID-code, with $\rho_i = D_i = \proj{\psi_i}$.

  One can obtain a set of such vectors also by turning the probability
  distributions on $n$ bits form~\cite{AhlswedeDueck:ID.A} into
  superpositions -- cf.~\cite{winter:q-ID-1} for details.
\end{example}

Fingerprinting ID-codes use quantum superpositions in a nontrivial
way, albeit the almost-orthogonality is somewhat analogous to the way 
the classical distributions in~\cite{AhlswedeDueck:ID.A} do not overlap too
much. However, they only use pure states, whereas the power of classical
identification comes from randomization. Hence it is natural to ask whether
mixed states offer any improvement. 
As the classical capacity of a noiseless qubit channel is $1$, the 
following result came as a bit of a surprise. It was proved using powerful 
geometric measure concentration techniques -- cf.~\cite{HLW:generic,ASW:Hastings} 
for other applications in quantum information theory.

\begin{theorem}[Winter~\cite{winter:q-ID-1}]
  \label{thm:ID-capacity-qubit}
  For the noiseless qubit channel $\id_2 = \id_{\CC^2}$, and 
  $0 < \lambda_1,\lambda_2, \lambda_1+\lambda_2 < 1$,
  \[
    2^{\Omega(2^{2n})} \leq N(n,\lambda_1,\lambda_2) \leq 2^{O(2^{2n})}.
  \]
  As a consequence, $C_{\rm ID}(\id_2) = 2$ and the strong converse holds.
  If the encodings are restricted to pure states, the capacity is only $1$.
  \qed
\end{theorem}
[In~\cite{winter:q-ID-1} (Remark 13; the technical argument there has been
elaborated in~\cite{Dupuis-et-al}) it was heuristically argued that
one would expect $C_{\rm ID}^{\rm sim}(\id_2)$ to be $1$ rather than 
$2$.\footnote{\textcolor{blue}{This conjecture has been proved recently~\cite{TSA:covering},
indeed by showing that $C_{\rm ID}^{\rm sim}(\id_2)$ is attained by codes 
using only pure states as encodings, so the above theorem applies. In fact,
the same argument proves that $C_{\rm ID}^{\rm sim}(\cN)$ for any channel 
is attained by codes using only pure state encodings~\cite{CDBW:0-sim-ID}.}}]

To appreciate why this result was so surprising, we need to go back to 
the insights from the original identification 
papers~\cite{AhlswedeDueck:ID.A,AhlswedeDueck:ID.B}: It was understood that
what determines identification capacity of a communication system is its
ability to establish common randomness (cf.~\cite{Ahlswede:GIT}),
as long as some sublinear amount of actual communication is available.
But the common randomness capacity of a noiseless qubit channel is $1$.
However, a noiseless qubit channel can also establish entanglement (ebits) 
at rate $1$. 
And indeed, in~\cite{winter:q-ID-2,remote-state-preparation} it was found
that $k$ EPR pairs shared between sender and receiver, together with $o(n)$ 
bits of communication are sufficient to identify $2^{\Omega(2^{2k})}$ messages.
In this respect, it may be interesting to draw attention to the following:

\begin{proposition}[Winter~\cite{winter:q-ID-1}]
  \label{prop:concatenation-w-randomness}
  Given an ID-code of rate $C$ and common randomness of rate $R$, one can
  construct an ID-code of rate $C+R-o(1)$ which uses the signal states
  of the first code and correlations with the common randomness.
  \qed
\end{proposition}

In other words: Whatever your communication system, its identification
capacity is increased by $1$ by each bit of common randomness. This was
used in~\cite{winter:q-ID-1} to derive a lower bound on the ID-capacity 
of a quantum channel: If $\cN$ permits simultaneous transmission of classical
bits and qubits at rates $C$ and $Q$, respectively, then
$C_{\rm ID}(\cN) \geq C+2Q$. Thus the results of~\cite{DevetakShor} become
applicable, where the $Q$-$C$ capacity region was determined. 
As we saw above, this bound, can be strictly larger
than the classical capacity $C(\cN)$ of the channel, marking a decisive 
departure from the behaviour of classical channels.

Beyond these bounds and a few special examples in~\cite{winter:q-ID-1},
the ID-capacity (simultaneous or not) of a general quantum channel 
remains elusive. However, in section~\ref{sec:Q-to-C} below we shall present 
a new lower bound.


\section{How to Identify Quantum States?}
\label{sec:q-ID}
So far the only quantum element in the discussion pertained to the channel
model. However, there is a natural way in which even the task of
identification can be extended from classical to quantum information.
This has been promoted in~\cite{winter:q-ID-1} and further in the more
recent~\cite{HaydenWinter:weak-decoupling}. In the following, $\cP(A) \subset \cS(A)$
denotes the set of pure quantum states on a system $A$.

\begin{definition}[Winter~\cite{winter:q-ID-1}]
  \label{defi:q-ID-code}
  A \emph{quantum ID-code for the channel $\cN$ with error $\epsilon$}, 
  for the Hilbert space $K$, is a pair of maps $\cE : \cP(K) \longrightarrow \cS(A)$ and 
  $\cD : \cP(K) \longrightarrow \cL(B)$ with $0\leq \cD_\varphi \leq \1$ for
  all $\varphi = \proj{\varphi}\in\cP(K)$, 
  such that for all pure states/rank-one projectors $\psi,\varphi\in\cP(K)$,
  \[
    \bigl| \tr\psi\varphi - \tr \cN\bigl(\cE(\psi)\bigr)\cD_\varphi \bigr| \leq \epsilon.
  \]
  If the encoding $\cE$ is cptp we speak of a \emph{blind} code, in general
  and to contrast it with the former, we call it \emph{visible}.
  
  For the case of an iid channel $\cN^{\ox n}$, we denote the maximum 
  dimension of a blind (visible) quantum ID-code by
  $M(n,\epsilon)$ ($M_v(n,\epsilon)$).
\end{definition}

This notion can be motivated as follows: In quantum transmission,
the objective for the receiver is to recover the state $\psi$ by means
of a suitable decoding (cptp) map $\widetilde{\cD}:\cL(B) \longrightarrow \cL(K)$,
with high accuracy.
Of course then the receiver can perform any measurement on the decoded
state, effectively simulating an arbitrary measurement on the original
input state, in the sense that for any state $\rho$ and POVM $M=(M_i)_i$
on $K$, there exists another POVM $M'=(M_i')_i$ on $B$ such that the
measurement statistics of $\rho$ under $M$ is approximately that of
$\cN(\cE(\rho))$ under $M'$.
($M'$ can be written down directly via the adjoint 
$\widetilde{\cD}^\dagger:\cL(K) \longrightarrow \cL(B)$ of the decoding
map, which maps measurement POVMs on $K$ to POVMs on $B$: 
$M_i' = \widetilde{\cD}^\dagger(M_i)$.)
The converse is also true: If the receiver can simulate sufficiently general
measurements on the input state by suitable measurements on the channel output, 
then he can actually decode the state by a cptp map $\widetilde{\cD}$~\cite{Renes10}.

This allows us to relax the task of quantum information transmission to
requiring only that the receiver be able to simulate the statistics of
certain restricted measurements. In the case of quantum identification,
these are $(\varphi,\1-\varphi)$ for arbitrary rank-one projectors 
$\varphi = \proj{\varphi} \in \cP(K)$. 
They are the measurements which allow the receiver to ask the (quantum)
question: ``Is the state equal to $\varphi$ or orthogonal to it?'' Obviously,
in quantum theory this question cannot be answered with certainty, but
for each test state it yields a characteristic distribution. The quantum-ID
task above is about reproducing this distribution. 

Note that we can always concatenate a blind or visible quantum ID-code
for the Hilbert space $K$ with a fingerprinting set of pure states in
$K$, to obtain a classical ID-code in the sense of 
Definition~\ref{defi:c-ID-code}. This is because in fingerprinting the 
encodings are pure states $\psi_i$ and the tests precisely the POVMs
$(\psi_i,\1-\psi_i)$.
Hence, as the cardinality of the fingerprinting set is exponential in
the dimension $|K|$, $M(n,\epsilon)$ and $M_v(n,\epsilon)$ can be at
most exponential in $n$.

\begin{definition}
  \label{defi:Q-ID-capacities}
  For a quantum channel $\cN$, the 
  \emph{blind, respectively visible, quantum ID-capacity} is defined as
  \begin{align*}
    Q_{\rm ID}(\cN)   &:= \inf_{\epsilon>0} 
                          \liminf_{n\rightarrow\infty} \frac{1}{n} \log M(n,\epsilon), \\
    Q_{\rm ID,v}(\cN) &:= \inf_{\epsilon>0} 
                          \liminf_{n\rightarrow\infty} \frac{1}{n} \log M_v(n,\epsilon).
  \end{align*}
  If we leave out the qualifier, the quantum ID-capacity is by default
  the blind variety.
\end{definition}
Note that by definition and the above remark,
\begin{equation}
  \label{eq:trivial-bounds}
  Q_{\rm ID}(\cN) \leq Q_{\rm ID,v}(\cN) \leq C_{\rm ID}(\cN).
\end{equation}

The first quantum ID-capacity that had been determined was for the
ideal qubit channel:

\begin{theorem}[Winter~\cite{winter:q-ID-1}]
  For the noiseless channel $\id_A$ on Hilbert space $A$, there exists a (blind)
  quantum ID-code with error $\epsilon$ and encoding a space $K$ of dimension 
  $|K| \geq C(\epsilon)|A|^2$, for some universal function $C(\epsilon) > 0$.
  
  As a consequence, $Q_{\rm ID}(\id_2) = Q_{\rm ID,v}(\id_2) = 2$, twice the
  quantum transmission capacity.
  \qed
\end{theorem}

In view of this theorem, we gain at least $2$ in capacity for each noiseless
qubit we use additionally to the given channel. This motivates the following
definition.

\begin{definition}[Hayden/Winter~\cite{HaydenWinter:weak-decoupling}]
  For a quantum channel $\cN$, the 
  \emph{amortized (blind/visible) quantum ID-capacity} is defined as
  \begin{align*}
    Q_{\rm ID}^{\rm am}(\cN)   &:= \sup_k\ Q_{\rm ID}(\cN \ox \id_k) - 2\log k, \\
    Q_{\rm ID,v}^{\rm am}(\cN) &:= \sup_k\ Q_{\rm ID,v}(\cN \ox \id_k) - 2\log k,
  \end{align*}
  respectively.
\end{definition}

The blind quantum ID-capacities are among the best understood, thanks to 
recently made conceptual progress, which we review in the next section.
We will then also ask the question \emph{how much} amortization is required.
This is formalized in the usual way: Namely, for a rate $Q \leq Q_{\rm ID}^{\rm am}(\cN)$,
we say that $A$ is an \emph{achievable amortization rate} if there exist 
$k_n$ for all $n$, such that
\[
  \liminf_{n\rightarrow\infty} 
      \frac{1}{n} \left( Q_{\rm ID}(\cN^{\ox n} \ox \id_{k_n}) - 2\log k_n \right) \geq Q
  \quad \text{and} \quad
  \limsup_{n\rightarrow\infty} 
      \frac{1}{n} \log k_n \leq A,
\]
giving rise to an achievable quantum ID-rate/amortization region, \emph{viz.}~a
tradeoff between $Q$ and $A$. Similarly of course for the visible variant.

\section{Weak Decoupling Duality}
\label{sec:weak-decoupling}
The fundamental insight about quantum information transmission, which allowed
an understanding of the quantum capacity as we have it today, is the
\emph{decoupling principle}: for a channel to permit (approximate) error
correction it is necessary and sufficient that it leaks (almost) no
information to the environment in the sense that the complementary
channel $\widehat{\cN}$ is close to constant. To be precise, $\id_{A'}\ox\widehat{\cN}$
should map an entangled test state $\Phi^{A'A}$ to $\approx \Phi^{A'} \ox \sigma^E$,
where the approximation is with respect to the trace norm on density operators.
In practice, to define capacities it is enough to demand this for the
maximally entangled test state between the code space and a reference 
system~\cite{SchumacherWestmoreland:decoupling}.

This condition is compactly expressed as saying that $\widehat{\cN}$ is
$\approx [\sigma^E]$ in the so-called diamond norm, the completely bounded
version of the naive super-operator norm. Here, $[\sigma^E]$ denotes the 
constant channel mapping every input to $\sigma^E$. Because of this
connection to completely bounded norms, we call channels with the above
property \emph{completely forgetful} or \emph{decoupling}.

Indeed, it is well-known that this is a much stronger condition than 
$\widehat{\cN}(\rho) \approx \sigma^E$ for all input states $\rho$ on $A$.
Cf.~\cite{HLSW:random} for some instances of this effect relevant to
quantum information processing. There, it is shown how to construct
channels that are only (approximately) \emph{forgetful} (or \emph{weakly
decoupling}), but far from completely forgetful.

To state the following conceptual points about blind(!) quantum ID-codes, 
it is useful to fix an encoding cptp map $\cE: \cL(K) \longrightarrow \cL(A)$
and to combine it with the noisy channel, $\cN' = \cN \circ \cE$, for
which we choose a Stinespring dilation $V:K \hookrightarrow B \ox F$.
The quantum ID-code is now the entire input space $K$ of this effective
new channel, together with the previously given operators $D_\varphi$
on $B$. The next result states that just as quantum error correctability
of $\cN'$ is equivalent to $\widehat{\cN'}$ being 
decoupling~\cite{SchumacherWestmoreland:decoupling,Kretschmann-et-al},
quantum identification is essentially equivalent to weak decoupling 
from the environment:

\begin{theorem}[Hayden/Winter~\cite{HaydenWinter:weak-decoupling}]
  \label{thm:ID-implies-weak-decoupling}
  If $K$ is a $\epsilon$-quantum ID-code for the channel $\cN'$
  with Stinespring dilation $V:K \hookrightarrow B \ox F$, then the
  complementary channel $\widehat{\cN'}$ is approximately forgetful:
  \[
      \forall \ket{\varphi},\ket{\psi} \in K\quad
           \frac{1}{2}\left\| \widehat{\cN'}(\varphi)-\widehat{\cN'}(\psi) \right\|_1
                          \leq \delta:=7\sqrt[4]{\epsilon}.
  \]
  
  Conversely, if $\widehat{\cN'}$ is approximately forgetful with error $\delta$,
  then the trace-norm geometry is approximately preserved by $\cN'$:
  \[
      \forall\ket{\varphi},\ket{\psi} \in S\quad
           0 \leq \big\| \varphi - \psi \big\|_1 - \big\| \cN'(\varphi) - \cN'(\psi) \big\|_1
                          \leq \epsilon := 4\sqrt{2\delta}.
  \]
  If, in addition, the nonzero eigenvalues of the environment's 
  states $\widehat{\cN'}(\varphi)$ lie in the
  interval $[\mu,\lambda]$ for all $\ket{\varphi} \in K$, then one can construct
  an $\eta$-quantum ID-code for $\cN'$ (i.e.~a set of operators $D_\varphi$ for
  all $\ket{\varphi} \in K$ as in Definition~\ref{defi:q-ID-code}), 
  with $\eta := 7\delta^{1/8}\sqrt{\lambda/\mu}$.
  \qed
\end{theorem}

\begin{remark}
  While it would be desirable to eliminate the eigenvalue condition at
  the end of the theorem, the condition is fairly natural in this
  context. If the environment's states $\widehat{\cN'}(\varphi)$ are very 
  close to a single state $\sigma^F$ for all $\ket{\varphi} \in K$, 
  then all the $V\ket{\varphi}$ are
  very close to being purifications of $\sigma^F$, meaning that they
  differ from one another only by a unitary plus a small perturbation.
  If $\sigma^F$ is the maximally mixed state or close to it, then the
  assumption will be satisfied. In the asymptotic iid setting we are
  looking at this turns to be the case.
\end{remark}

This characterization of quantum ID-codes (albeit ``only'' blind ones)
allows the determination of capacities by a random coding argument,
for which only the weak decoupling has to be verified. The above duality
theorem is not only the basis for the direct but also for the converse
part(s) of the following capacity theorem.

\begin{theorem}[Hayden/Winter~\cite{HaydenWinter:weak-decoupling}]
  \label{thm:Q-ID}
  For a quantum channel $\cN$, its (blind) quantum ID-capacity is given by 
  \[\begin{split}
    Q_{\rm ID}(\cN)       &= \lim_{n\rightarrow\infty} 
                                     \frac{1}{n} Q_{\rm ID}^{(1)}(\cN^{\ox n}), \text{ where} \\
    Q_{\rm ID}^{(1)}(\cN) &= \sup_{\ket{\phi}}
                                     \bigl\{ I(A:B)_\rho \text{ s.t. } I(A \rangle B)_\rho > 0 \bigr\},
  \end{split}\]
  where $\ket{\phi}$ is the purification of an input state to $\cN$,
  $\rho^{AB} = (\id\ox\cN)\phi$ and $I(A:B)_\rho = S(\rho^A) + S(\rho^B) - S(\rho^{AB})$
  is the mutual information, and 
  $I(A \rangle B)_\rho = S(\rho^B) - S(\rho^{AB})$ the coherent information (which already
  appeared in eq.~(\ref{eq:Q-cap})). 
  We declare the $\sup$ to be $0$ if the set above is empty.
  In particular, $Q_{\rm ID}(\cN) = 0$ if and only if $Q(\cN) = 0$.
  
  Furthermore, the amortized quantum ID-capacity equals
  \[
    Q_{\rm ID}^{\rm am}(\cN) = \sup_{\ket{\phi}} I(A:B)_\rho = C_E(\cN),
  \]
  the entanglement-assisted classical capacity of $\cN$~\cite{BSST}.
  \qed
\end{theorem}

\begin{remark}
  \label{rem:sufficiently-low-noise}
  Let us say that a channel $\cN$ has ``sufficiently low noise'' if for an input 
  state $\ket{\phi}$ maximizing $I(A:B)_\rho$, $\rho = (\id\ox\cN)\phi$,
  it holds that $I(A \rangle B)_\rho > 0$. 
  This is motivated by the fact that in this case the channel has positive 
  quantum capacity. Also, for any channel, $\cN\ox\id_k$ has sufficiently
  low noise if $k$ is chosen large enough; likewise $p\cN + (1-p)\id$ if
  $p>0$ is small enough.
  
  In that case, the above tells us
  $Q_{\rm ID(\cN)} = Q_{\rm ID}^{\rm am}(\cN) = \sup_{\ket{\phi}} I(A:B)_\rho$,
  which is an additive, single-letter formula.
\end{remark}

This theorem also shows that amortized and non-amortized quantum ID-capacities are 
different -- indeed, any channel $\cN$ with vanishing quantum capacity also 
has $Q_{\rm ID}(\cN) = 0$, whereas $Q_{\rm ID}^{\rm am}(\cN) = 0$ only for 
trivial channels. In particular this implies that $Q_{\rm ID}$ is not
additive.
In~\cite{HaydenWinter:weak-decoupling} it is in fact proven that certain channels
require a positive rate of amortization to attain or even to approximate
$Q_{\rm ID}^{\rm am}$. The example analyzed there is the qubit erasure channel
\begin{align*}
  \cE_q : \cL(\CC^2) &\longrightarrow \cL(\CC^3) \\
          \rho       &\longmapsto     (1-q)\rho \oplus q\proj{\ast},
\end{align*}
which will serve us again in the following section. To be precise, for
$0\leq q < \frac12$, the channel has sufficiently low noise and no
amortization is required. For $\frac12 \leq q \leq 1$ instead, an amortized
rate of at least $2q-1$ qubits per channel use are necessary.

\medskip
On the other hand, for all \emph{symmetric} channels, i.e.~those with
$E=B$ in the Stinespring representation and $\cN = \widetilde{\cN}$,
whereas quantum capacity and hence $Q_{\rm ID}$ are zero, only a vanishing 
rate of amortization is necessary to attain $Q_{\rm ID}^{\rm am}$. This
is because they have $I(A \rangle B)_\rho = 0$ for every input state, so
arbitrarily little is required to make the coherent information positive.

This includes qc-channels with rank-one POVM $(M_y)_{y\in\cY}$, and
the noiseless classical bit channel
\[
  \overline{\id}_2 : \rho \longmapsto \sum_{b=0,1} \proj{b} \rho \proj{b}.
\]
The latter implies that also cq-channels $\cN$ only require a vanishing rate of
amortization to attain $Q_{\rm ID}^{\rm am}(\cN) = C(\cN) = C^{(1)}(\cN)$:
This is because we can use $n\gg 1$ copies of $\cN$, with appropriate encoding
and decoding, to simulate $\bigl(C(\cN)-o(1)\bigr)n$ almost noiseless classical
bits. This also shows that the rate $C(\cN)$ is attainable for all channels
$\cN$ as an amortized quantum ID-rate, with vanishing rate of amortization.

In fact, inspection of the proof of the direct part of Theorem~\ref{thm:Q-ID}
(Thm.~12 in~\cite{HaydenWinter:weak-decoupling}) reveals that for the noiseless
classical channel $\overline{\id}_2$, a constant amount of amortization is
enough, hence the same for all cq-channels, and also for certain rank-one
POVM qc-channels, namely those for which the outputs $\cN(\tau)$ and
$\widetilde{\cN}(\tau)$ for the maximally mixed input state $\tau_A$
are themselves maximally mixed. Because then the typicality arguments in
the proof, which deal with eigenvalue fluctuations around the inverse
exponential of the entropy, are unnecessary.

\begin{remark}
  \label{rem:visible-cq-channel}
  The previous observations show that the amortized quantum ID-capacity of a
  cq-channel $\cN$ (which equals its classical capacity)
  can be achieved by visible, non-amortized codes:
  \[
    Q_{\rm ID,v}(\cN) = Q_{\rm ID}^{\rm am}(\cN) = C(\cN) = C^{(1)}(\cN).
  \]
  Indeed, choose a sequence of amortized quantum ID-codes for $n$ 
  uses of the channel, with amortized noiseless communication of a 
  system of dimension $t = o(\log n)$.
  Then, whatever the code produces as the input state $\omega = \cE(\psi)$ 
  to the channel $\cN^{\ox n} \ox \id_t$, the effect is the same if we first
  dephase the input to $\cN^{\ox n}$ as the channel is cq, so w.l.o.g.
  \[
    \omega = \sum_{x^n} p_{x^n} \proj{x^n} \ox \omega_{x^n},
  \]
  with states $\omega_{x^n} \in \cS(\CC^t)$. The latter can be
  described classically to good approximation using $o(n)$ 
  bits~\cite{HLW:generic,winter:q-ID-1}, which
  can be communicated by $o(n)$ uses of the channel (if we exclude the
  trivial case of zero capacity).
  
  This then is the visible scheme: the encoding of state $\psi$ is to
  sample from the distribution $p_{x^n}$ and sending $\proj{x^n}$ through
  $\cN^{\ox n}$, and to send a classical description of $\omega{x^n}$
  via $\cN^{\ox o(n)}$. The receiver creates then $\omega_{x^n}$ in 
  addition to the other channel output, and otherwise uses the measurement
  $D_\varphi$ from the amortized ID-code.
  
  That the capacity cannot be larger than $C(\cN)$ follows from 
  eq.~(\ref{eq:trivial-bounds}) and Theorem~\ref{thm:quantum-channel-ID}.

  On the other hand, $Q_{\rm ID}(\cN) = 0$ by Theorem~\ref{thm:Q-ID},
  so we obtain a separation between blind and visible quantum ID-capacity,
  a question left open in~\cite{winter:q-ID-1}.
  \qed
\end{remark}

\medskip
We close this section by pointing out that $Q_{\rm ID}^{\rm am}$ is one 
of only two fully understood identification capacities so far: it 
has a single letter formula which can be evaluated efficiently and it is additive.
The other one is the classical ID-capacity of a quantum channel with
``coherent feedback'' (meaning that in each use of the channel, the
environment of the Stinespring isometry ends up with the sender), which
we did not discuss here; the interested reader is referred to~\cite{winter:q-ID-2}.

\section{From $Q_{\rm ID}$ to $C_{\rm ID}$}
\label{sec:Q-to-C}
As pointed out in section~\ref{sec:q-ID}, concatenating a quantum
ID-code (blind or visible) with the fingerprinting construction
(Example~\ref{ex:fingerprinting}),
yields a classical ID-code of asymptotically the same rate. Hence,
\[\begin{split}
  C_{\rm ID}(\cN)          &\geq Q_{\rm ID,v}(\cN) \geq \begin{cases}
                                                          Q_{\rm ID}(\cN), \\
                                                          C(\cN),
                                                        \end{cases} \\
  C_{\rm ID}^{\rm am}(\cN) &\geq Q_{\rm ID}^{\rm am}(\cN) = C_E(\cN),
\end{split}\]
where the amortized classical ID-capacity is defined analogously to
the quantum variant.

Perhaps we do not find the amortized classical ID-capacity that interesting,
but at least we get lots of channels for which $C_{\rm ID}(\cN) \geq C_E(\cN)$,
namely all sufficiently low noise channels and of course all cq-channels.
This bound improves on the earlier best bound 
\[
  C_{\rm ID}(\cN) \geq \max \{ C+2Q : (Q,C) \text{ jointly achievable} \},
\]
the right hand side of which is always $\leq C_E(\cN)$. 
For example for the erasure channel $\cE_q$, the quantum-classical-capacity 
region is known~\cite{DevetakShor} to be
\[
  \text{conv}\left\{ (0,0),\ (0,1-q),\ \bigl((1-2q)_+,0\bigr) \right\},
\]
so the above maximization yields
\[
  C_{\rm ID}(\cE_q) \geq \begin{cases}
                           2-4q & \text{ for } 0 \leq q \leq \frac13, \\
                           1-q  & \text{ for } \frac13 \leq q \leq 1.
                         \end{cases}
\]
Our new bound instead is
\[
  C_{\rm ID}(\cE_q) \geq \begin{cases}
                           2-2q & \text{ for } 0 \leq q < \frac12, \\
                           1-q  & \text{ for } \frac12 \leq q \leq 1,
                         \end{cases}
\]
which is strictly better in the interval $[0,\frac12)$.

\section{Conclusion and Open Questions}
\label{sec:conclusion}
As it should have become clear from the above exposition, identification
theory in the quantum setting is an enormously fruitful area, much more so
even than the classical version, if only because we have at least five
natural capacities. And we did not yet even touch upon a general theory of
information transfer in quantum information, or rather how quantum information 
would fit into this far-reaching vision~\cite{Ahlswede:GTIT,Ahlswede:GTIT-updated},
these aspects still awaiting development.

At the same time the subject of identification via quantum channels is wide 
open, with most of the questions implied in the original 
papers~\cite{Loeber:PhD,AhlswedeWinter:ID-q,winter:q-ID-1} remaining
unsolved, despite significant progress over the last decade. In particular,
it turned out that the quantum identification task lent itself much more
easily to the currently available techniques, and that the recent progress
satisfyingly shed a fresh light on the older, and seemingly more elementary
classical identification task.
The following seven broad open problems are recommended to the reader's attention.

\begin{enumerate}
  \item Surely the biggest open problem is to determine the classical ID-capacity
    $C_{\rm ID}(\cN)$ of a general quantum channel, and to study its properties,
    such as additivity etc.
    Even obtaining non-trivial upper bounds would be a worthy goal. Note that 
    practically all transmission capacities of a channel are upper bounded by
    its entanglement-assisted capacity, by way of the Quantum Reverse Shannon 
    Theorem~\cite{BSST,QRST,Berta-et-al} through simulation of the channel by
    noiseless communication and unlimited shared entanglement. This argument is
    not available here since entanglement or even common randomness have an
    impact on the ID-capacities.
    
    In fact, the few cases for which $C_{\rm ID}$ is known are consistent
    with the idea that it is always equal to the entanglement-assisted classical
    capacity of the channel~\cite{winter:q-ID-1}. One might speculate that 
    $C_{\rm ID}(\cN) \geq Q_{\rm ID,v}(\cN) \geq C_E(\cN)$ be true for 
    all channels, seeing that for sufficiently low noise we can prove it,
    and that it is true for the \emph{amortized} classical ID-capacity.
    The erasure channel $\cE_q$ discussed in section~\ref{sec:Q-to-C} is already
    an excellent test case for this idea.

  \item Is there a deeper, operational, reason why the amortized quantum ID-capacity
    equals the entanglement-assisted classical(!) capacity of a channel? In
    the derivation of~\cite{HaydenWinter:weak-decoupling} this comes 
    out naturally as a result of the analysis, but almost as an accident, and
    it seems difficult to connect it to~\cite{BSST}...

  \item Is the simultaneous ID-capacity $C_{\rm ID}^{\rm sim}(\cN)$ equal
    to the non-simultaneous version $C_{\rm ID}(\cN)$? 
    I suspect that they are different, possibly even
    for the noiseless qubit channel (see Theorem~\ref{thm:ID-capacity-qubit}
    and subsequent remarks). In such a case we face another problem to
    determine $C_{\rm ID}^{\rm sim}(\cN)$.
    When studying simultaneous ID-codes, L\"ober's technical condition in
    Theorem~\ref{thm:quantum-channel-ID} deserves special attention, as
    it precludes using the entire input state space of the iid 
    channels.\footnote{\textcolor{blue}{Cf.~footnote 1 for the recent proof 
    that indeed $C_{\rm ID}^{\rm sim} \neq C_{\rm ID}$~\cite{TSA:covering}.
    The arguments in \cite{TSA:covering,CDBW:0-sim-ID} allow the determination 
    of the simultaneous ID-capacity for some channels, but leave the general
    case wide open.}}
    A very interesting case to study will be (rank-one POVM) qc-channels as
    there any identification code is \emph{per se} simultaneous. For 
    these channels we know the amortized quantum ID-capacity (it is the 
    entanglement-assisted classical capacity, which evaluates to $\log |A|$), and
    that amortization rate $0$ is sufficient to achieve it, in some cases
    even a constant amount. In fact, it would be interesting to know
    whether the visible quantum ID-capacity for these channels is the
    same -- cf.~the case of cq-channels discussed in 
    Remark~\ref{rem:visible-cq-channel} --; this would evidently prove
    ${\color{blue} C_{\rm ID}^{\rm sim}(\cN) =}\ C_{\rm ID}(\cN) \geq \log |A|$ 
    for all these channels $\cN$, whereas
    it is known that the classical capacity $C(\cN)$ for many of them is
    much smaller.

  \item The role of amortization is extremely interesting: For the quantum
    ID-capacity it makes for a quasi-superactivation effect, since a vanishing
    rate of it (i.e.~an arbitrary small rate of noiseless communication)
    can turn a capacity $0$ channel into one of positive capacity. 
    It is possible that vanishing rate of amortization likewise has an
    impact on classical ID-capacities -- see the example of qc-channels discussed
    in the previous point.
    
    Finally, in~\cite{HaydenWinter:weak-decoupling} only the non-triviality
    of amortization (and only for the erasure channels) was proved. 
    How to characterize the quantum ID-rate vs.~amortization rate tradeoff?

  \item We have seen that the visible quantum ID-capacity can be larger than 
    the blind variant, indeed the former can be positive while the latter is
    $0$ for cq-channels.
    Let us note that the distinction visible/blind can also be made in the
    quantum transmission game, and there it is far from clear whether
    there will be a difference in capacities, see~\cite{winter:q-ID-1}.

  \item We did not comment much on the role of shared correlations in the
    identification game, indeed referring the reader to~\cite{winter:q-ID-1,winter:q-ID-2},
    where also the impact of feedback is discussed.
    However, in Proposition~\ref{prop:concatenation-w-randomness} we saw that
    not only is the classical ID-capacity of common randomness (in the presence
    of negligible communication) equal to $1$ per bit, but it increases the rate of any
    given ID-code by $1$ per bit. We also know that the classical ID-capacity of shared
    entanglement is $2$ per ebit, but it is open whether we can augment a
    given ID-code with entanglement to increase its rate by $2$ per ebit.

  \item Finally: All the known upper bounds on classical ID-capacities are in fact
    strong converses. Does the strong converse also hold for 
    (visible, blind, amortized, etc) quantum ID-capacities? This question seems
    to require new techniques to be answered.

\end{enumerate}

\section*{Acknowledgements}
I have thought about identification in quantum information theory for quite 
some time, going back all the way to the days of my PhD, from which period date
my hugely enjoyable mathematical interactions with Peter L\"ober, and of course with Rudolf 
Ahlswede, a sometimes terrifying but always, and ever newly, inspiring \emph{Doktorvater}.
In an article such as the present one it may be permitted to say that I miss him. 
Why, he keeps influencing my work even now!


Fortunately, later on others began to share my enthusiasm for quantum 
identification, most importantly Patrick Hayden. Much of the present review 
was only motivated by our recent collaboration.

I am or have been supported by a U.K.~EPSRC Advanced Research 
Fellowship, the European Commission (STREP ``QCS'' and
Integrated Project ``QESSENCE''), the ERC Advanced Grant ``IRQUAT'',
a Royal Society Wolfson Merit Award and a Philip Leverhulme Prize.

\bibliographystyle{splncs}

\end{document}